# DEFINITION OF MULTISPECTRAL CAMERA SYSTEM PARAMETERS TO MODEL THE ASTEROID 2001 SN263

Gabriela de Carvalho Assis Goulart[a], Thiago Statella[b,*] and Rafael Sfair[c, d]

[a] Universidade Federal de Mato Grosso, Av. Fernando Corrêa da Costa, 2367 Boa Esperança, 78060-900, Brazil
e-mail: gabrielagoulart.eng@gmail.com
[b] Instituto Federal de Educação, Ciência e Tecnologia de São Paulo, 27-50 José Ramos Júnior, 19470-000, Brazil
e-mail: t.statella@gmail.com
[c] São Paulo State University (UNESP), School of Engineering and Sciences, Guaratinguetá, 12516-410, Brazil
e-mail: rafael.sfair@unesp.br
[d] Institut für Astronomie und Astrophysik, Universität Tübingen, Auf der Morgenstelle 10, 72076 Tübingen, Germany
[*] Correspondence: gabrielagoulart.eng@gmail.com

**Abstract**

In 2012, Brazil began the studies to send its first deep space exploration mission, ASTER, which would be the first mission to orbit a triple asteroid system, 2001 SN263. We aim to contribute to the ASTER mission by defining the parameters of a multispectral camera system that will be used to study the asteroid system 2001 SN263, through software simulations that should help planning the data collection. We inserted the shape model of the objects in the software POV-Ray and modeled two cameras, a Wide Angle (WAC) and a Narrow Angle (NAC). We inserted the asteroid's parameters and simulated the satellite position. We created various scenes so we could obtain a good view of the asteroid. Alpha is entirely visible only in the WAC images, while the NAC is expected to reveal surface details. Beta seems relatively small in the WAC images, whereas we obtain a broad view from the NAC at 100 km distance. Gamma, smaller than Beta, should provide more detailed images through the NAC, whereas the WAC images should be able to show its inclined orbit around Alpha. To see Gamma behind Alpha in its revolution movement, we would have to elevate the camera's orbit. The method employed to simulate images generated by satellite cameras can be applied to other scenarios where the target requires imaging, extending beyond the field of planetary geology.

## 1. Introduction

Asteroids can tell us the history of our solar system, since they are leftover debris from the accretion of planetesimals and come directly from the primitive solar nebula (Ryan, 2000). Therefore, they may provide samples suitable for dating the formation of the Solar System (Vita-Finzi, 2013). The distribution, formation, evolution, and physical nature of asteroids are key elements to understand planet formation and why there is life on Earth. Through the investigation of the formation and evolution of asteroids, we learn more about the origins and characteristics of debris disks and planetary systems around other stars (Michel et al., 2015). Scientists can limit, for example, the potential range of initial conditions that might lead to the formation of a planetary system resembling our own, by studying the orbital and physical features of the asteroid belt and the impact craters on terrestrial planets caused by asteroids (Bottke et al., 2002). Asteroids have also been studied in terms of their significance both as potential hazards and valuable resources for Earth. While the chances of Earth being impacted by an asteroid in the near future are quite remote, the potential outcomes of such an event could be profoundly significant for life on our planet (Michel et al., 2015). Most recently, commercial exploration is being considered as well (Andrews et al., 2015, Mazanek et al., 2015, Bonin et al., 2016), since asteroids contain rare-Earth elements, including metals of value, and water or water components useful for space travel (Michel et al., 2015).

Asteroids can also play an important role in creating the conditions for habitability on Earth-sized planets. Meteorites contain water, organics, and amino acids that are potential precursors for life and provide insights into the composition of terrestrial planets and may have contributed to the formation of oceans (Michel et al., 2015). According to Vita-Finzi (2013), organic molecules may have been imported by comets, interplanetary dust particles, and carbonaceous asteroids small enough to be significantly decelerated and large enough to penetrate the atmosphere. If, in the past, we wondered whether asteroids had satellites (Sullivan et al., 2002), today the questions revolve around how the dynamics between the main object and its satellites work. Asteroids with satellites are excellent systems for studies on internal structure and composition (Santana-Ros et al., 2017).

Electromagnetic radiation is one of the main energy forms that we can passively observe (McLean, 2008). Common methods to study asteroid composition include using light curves and spectral observations, both of which can be conducted with ground-based telescopes. Studies of asteroid light curves provide information on vital properties such as rotation rates, shape, pole orientation, and surface characteristics (Emery et al., 2015). However, light curves tend to be biased against small, slowly rotating bodies, because the weak luminosity of these objects results in a skewed representation in the light curves, thereby limiting opportunities for a detailed study of these features (Benner et al., 2015). Reflectance spectra, ranging from UV to mid-infrared wavelengths, along with emission features in the mid-infrared range, are utilized to identify minerals and other compounds on the surfaces of asteroids or meteorites (DeMeo et al., 2015), but although taxonomy is useful for grouping objects with similar spectral shapes, it does not always diagnose the mineralogy and composition of celestial objects, and visible spectra alone cannot determine mineralogy (Dementieva & Ostrogorsky, 2012). Moreover, for ground-based observatories, light must traverse the atmosphere, which not only degrades image quality due to turbulence but also results in poor or non-transmission of certain electromagnetic spectrum segments including most of the infrared and all of the UV, X-ray, and gamma-ray regions (McLean, 2008). Therefore, launching a spacecraft offers the advantage of conducting direct measurements, including the observation of surface features like craters, and allows for a more precise analysis of specific features thanks to the improved spatial resolution (or Instantaneous Field-of-View - IFOV) of the spacecraft data (Reddy et al., 2015). When studying a system with multiple components, such as 2001 SN263, a space mission is even more beneficial because these systems provide a natural laboratory for exploring various physical processes that affect asteroids and for examining how their dynamics can offer valuable insights into their physical properties (Michel et al., 2015). Thus, space missions play a crucial role in the study of asteroids.

To better study objects that orbit close to Earth (near-Earth objects; NEOs), in 2012 Brazil began the studies to send its first deep space exploration mission, called ASTER, which would be the first mission to orbit a triple asteroid system (Macau et al., 2012), the (153591) 2001 SN263. During the mission, the spacecraft should orbit the triple-system asteroid 2001 SN263 from 100 km, in the reconnaissance position, and then navigate as close as 30 km from the target, in the mapping position, from where the primary objectives of the mission should be fulfilled. In addition, there could be an extended phase, where the spacecraft would get closer to the system and perform a small-scale mapping of previously selected features in the reconnaissance and mapping positions.

In its payload, the platform would carry a suite of three scientific instruments: a laser altimeter, named Aster Laser Rangefinder (ALR), a spectrometer, named Aster Mission Infrared Spectrometer (AMIS), and a camera system, the Aster Multispectral Cameras (AMC), operating in the visible and near-infrared spectrum.

The ALR should provide information about distance of the spacecraft to the target and relative velocity between them, mainly in the approximation stages. ALR can also be used to produce a topographic profile of the system, and by combining ALR data with data from the imaging cameras, it should be possible to characterize geodesic and geophysical properties of the three bodies. Finally, ALR will be used to calibrate the infrared spectrometer (Brum et al., 2011).

The camera system should provide a complete coverage of the three targets, namely, Alpha, Beta and Gamma (with diameters of 3.2, 1.2 and 0.7 km, respectively (Becker et al., 2015). The snapshot capability of the camera is essential to produce a fast initial mapping of the targets as soon as the spacecraft reaches the reconnaissance orbit, providing support for the other instruments onboard. The camera system will be used for shape and size definition, rotation monitoring, limb pointing for shape model construction and surface photometry, and is the main instrument to provide the complete coverage of the 2001 SN263.

In this work, we aim to contribute to the ASTER mission by defining the parameters of a multispectral camera system that could be used to study the asteroid system 2001 SN263, through software simulations that should help planning the data collection. We also aim to create a cartographic projection of the three bodies.

The paper is structured as follows. In Section 2 we present the characteristics of the asteroid (153591) 2001 SN263. In Section 3 we present the scientific heritage from similar missions. In Section 4 we present how we defined the parameters of the system utilizing a ray-tracing software. In Section 5 we present our results and discussion. Finally, in Section 6, we offer a summary of the results presented in the preceding section and make final comments.

## 2. The Asteroid (153591) 2001 SN263

Asteroid 2001 SN263 is a near-Earth asteroid (NEA) discovered in 2001 by the Lincoln Near-Earth Asteroid Research (LINEAR) program. Further radar observations carried out in Arecibo in 2008 revealed that it is a system composed of three bodies, the first known triple system (Becker et al., 2015; Nolan et al., 2008; Stokes et al., 2000). The triplet is formed by a main, larger body named Alpha, which is orbited by the two other ones, Beta and Gamma. Their shapes have been modeled by Becker et al. (2015) and the orbital and physical parameters of the bodies are given in Table 1. Here, *a* and *b* are the equatorial principal axes of the ellipsoid whereas *c* is its polar axis.

**Table 1**. Alpha, Beta and Gamma models as estimated by Becker et al. (2015). Parameters *a* and *b* are the equatorial principal axes of the ellipsoid whereas *c* is its polar axis.

|  | Alpha | Beta | Gamma |
| --- | --- | --- | --- |
| Ellipsoid axes (m) | $a = 2800 \pm 100$<br>$b = 2700 \pm 100$<br>$c = 2900 \pm 300$ | $a = 750 \pm 120$<br>$b = 1000 \pm 200$<br>$c = 640 \pm 100$ | $a = 540 \pm 170$<br>$b = 420 \pm 130$<br>$c = 410 \pm 130$ |
| Rotation period (h) | $3.43 \pm 0.0002$ | $13.43 \pm 0.0010$ | $16.40 \pm 0.0400$ |
| Orbit semi-major axis (km) |  | $16.63 \pm 0.1630$ | $3.80 \pm 0.0020$ |
| Orbital period (h) |  | $149.40 \pm 2.2800$ | $16.46 \pm 0.0480$ |

According to Winter et al. (2020), the volumes of Alpha, Beta, and Gamma are 8.34 km³, 0.24 km³, and 0.04 km³, respectively.

In Figure 1, we show the models derived from the analysis of the light curves for Alpha, Beta and Gamma, respectively, from Becker et al. (2015).

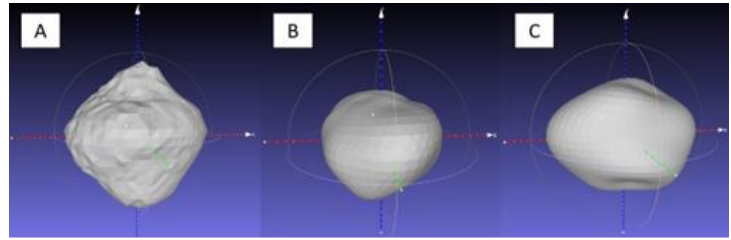

**Figure 1.** Alpha (A), Beta (B) and Gamma (C) tridimensional models, derived from light curve analysis (Becker et al., 2015). The axes X, Y and Z are presented in red, green and blue, respectively (not in relative scale).

In Figure 2, we show a representation of the system with the distances of the orbits in scale.

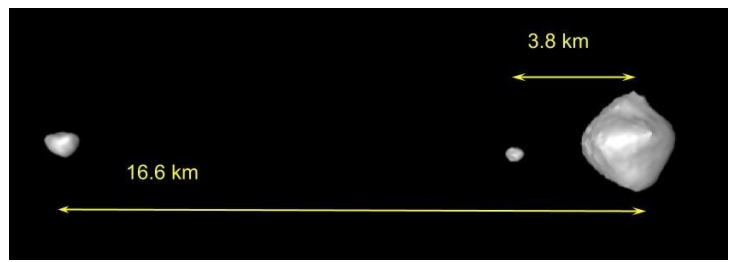

**Figure 2.** Distance of the Beta (farther) and Gamma (closer) satellites orbits from Alpha.

Regarding the spectral classification, it was first believed that the 2001 SN263 could be a C-type asteroid system, or at least fit in the C-group of carbonaceous asteroids

(Betzler et al., 2008). Later, Perna et al. (2014) concluded that the system is a B-type (C-group) ultra-blue asteroid, like Pallas and Bennu, its surface resembles those of heated CI carbonaceous chondrites, and it is probably very rich in organics, magnetite, and phyllosilicates.

The spectral signature of C-complex asteroids (according to the taxonomy developed by DeMeo et al., 2009), tend to have relatively featureless spectra (Burbine, 2002), which corresponds to carbonaceous meteorites. Higher-resolution spectral surveys (e.g., Bus & Binzel, 2002) have shown that almost half of the observed C-complex asteroids have a 0.7 μm feature (Burbine, 2002). The asteroid 2001 SN263, a C-complex and B-class asteroid (Ostrowski et al., 2011), has very low albedo (~0.06) and relatively flat spectrum as we can see in Figure 3.

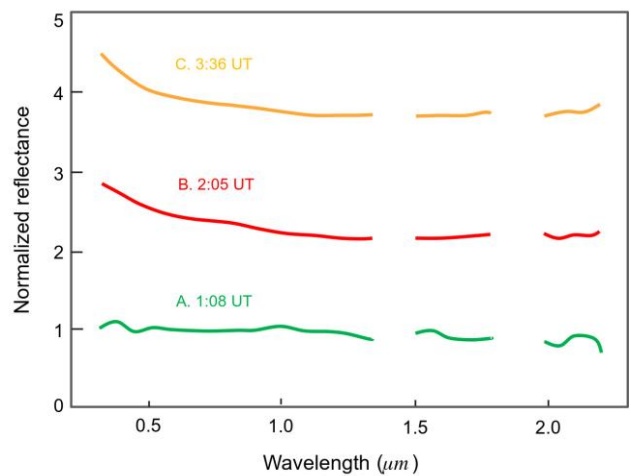

**Figure 3.** Asteroid 2001 SN263 spectra (the displacement is for visualization purpose) obtained on June 24, 2011, at three different times (after Perna et al., 2014).

A commonly used set of filters used to study asteroids is the Eight-Color Asteroid Survey (ECAS) filters, established by Tedesco et al. (1982). Considering the wavelength extent of the ECAS filters (*u*, *b*, *v*, *w*, *x*, *p* and *z* bands), the main spectral features we see in the reflectance curves in Figure 3 are found at the ultraviolet and near-infrared wavelengths. The *u* (235-415 nm) and *b* (355-560 nm) bands are sensitive to absorption in ferrous silicates, the *p* band (855-1055 nm) is located at the center of an absorption due to pyroxene and the *z* band (955-1155 nm) is located near the center of absorption by olivine. The primary diagnostic feature in olivine is a combined absorption feature at around ~1 μm, comprising three separate absorption bands located at 0.9 μm, 1.1 μm, and 1.25 μm, while pyroxenes display two absorption bands, one near ~1 μm and one at ~2 µm (Dementieva & Ostrogorsky, 2012). Figure 4 shows the reflectance of pyroxene, and we can see the absorption feature around 0.9 µm.

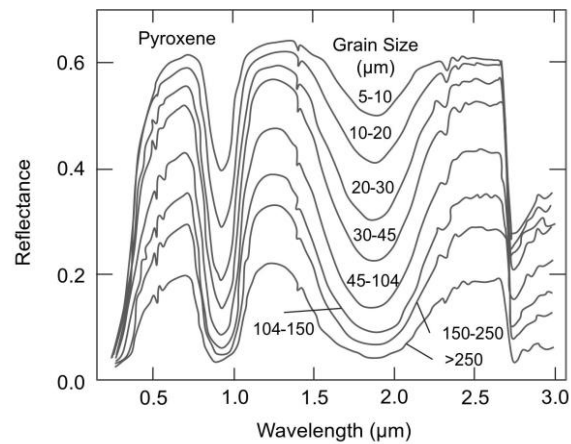

**Figure 4.** Pyroxene spectral signatures (Clark, 1993).

### 3. Scientific heritage from similar missions

Several deep-space missions have used optical remote sensing systems in the visible-infrared spectrum to study comets and asteroids. In 1991, the Galileo spacecraft orbited 951 Gaspra, producing 57 images of the asteroid with the Galileo Solid State Imaging (SSI) camera. The SSI had seven filters, not including the clear filter: violet, green, red, methane 1, methane 2, near-infrared and 1-µm. The spacecraft, whose destination was Jupiter, had also an ultraviolet spectrometer, a near-infrared mapping spectrometer and a photopolarimeter radiometer (Helfenstein et al., 1994; Galileo, 2021). Galileo also orbited the asteroid 243 Ida in 1993, and these were the first asteroid encounters in the history of spacecraft exploration of the solar system (Chapman, 1996).

NEAR Shoemaker was the first mission of NASA's new Discovery program. The spacecraft was equipped with a Multispectral Imager (MSI) camera. It encountered 253 Mathilde, a C-type asteroid, in 1997, when it was on its trajectory to 433 Eros, and the MSI camera could achieve a resolution of about 160 m/pixel at its closest approach, 1212 km (Veverka et al., 1999).

The NEAR Shoemaker mission to study the 433 Eros asteroid (~17 km) also carried a multispectral camera, with bands centered at 450, 550, 760, 900, 950, 1000, and 1050 nm plus one broad-band filter used for navigation. The bands were optimized to identify iron-containing silicate minerals. The field-of-view (FOV) of the camera was ~2.5° × 2.5° with ~25 × 25 µrad/pixel spatial resolution.

Hayabusa, the asteroid sample return mission operated by the Japan Aerospace Exploration Agency (JAXA), touched (25143) Itokawa (an S-type asteroid ~400 m in diameter) in 2005; the spacecraft was equipped with the Asteroid Multi-band Imaging Camera (AMICA), that had a wide band-pass filter and seven narrow band filters: 380 (*ul*), 430 (*b*), 550 (*v*), 700 (*w*), 860 (*x*), 960 (*p*), and 1010 nm (*zs*) (Yoshikawa et al., 2006). Seven of AMICA's filters were compatible with the Eight Color Asteroid Survey (ECAS), the standard for asteroid taxonomy in ground-based observations, and has an effective FOV of 5.83° × 5.69°, covered by 1024 × 1000 pixels, which corresponds to a spatial resolution of 0.7 m at the nominal distance of 7 km from Itokawa's surface (Ishiguro et al., 2010). The Hayabusa spacecraft also carried the Light Detection and Ranging (LIDAR), a laser altimeter responsible for measuring the distance from the target, guiding the probe at the initial encounter to Itokawa (Tsuno et al., 2017). During the mission phase, the spacecraft remained in a station-keeping heliocentric orbit, initially from ~20 km of the asteroid's surface (Ishiguro et al., 2010). Hayabusa2 used a similar narrow angle camera (ONC-T) plus two wide-angle cameras (ONC-W1 and ONC-W2) to study the asteroid Ryugu (~1 km in diameter, Sugita et al., 2013). The ONC-T FOV is 6.27° × 6.27°, and it has a wheel with seven bandpass filters and a panchromatic window: 390 nm (*ul*-band), 480 nm (*b*-band), 550 nm (*v*-band), 589 nm (*Na*), 700 nm (*w*-band), 860 nm (*x*-band), and 950 nm (*p*-band), based, except the *Na*, on the ECAS (Kameda et al., 2017). The ONC-W1 FOV is 69.71° × 69.71° and the ONC-W3 FOV is 68.89° × 68.89° (Suzuki et al., 2018). Like Hayabusa, the Hayabusa2 spacecraft also carried the LIDAR, mainly used as a navigation sensor for rendezvous and touchdown (Mizuno et al., 2017).

The Marco Polo mission to the near-Earth asteroids Ryugu and 2008 EV5 was never launched, however, according to its assessment study report (Marco Polo Science Study Team, 2009) the spacecraft would carry two cameras, a narrow-angle (~2° × 2° FOV) and a wide-angle (~11° × 11° FOV), with spectral bands matching the ECAS filters.

The Rosetta mission, whose final target was comet 67P/Churyumov-Gerasimenko (~4 km in diameter), visited asteroids (2867) Steins and (21) Lutetia in its trajectory, and determined their physical parameters, surface morphology, mineralogical composition, and searched for possible binary systems. (Keller et al., 2007). In 2014, Rosetta acquired images of the comet using a narrow angle camera (OSIRIS-NAC, ~2° × 2° FOV, ~19 × 19 μrad/pixel) and a wide angle camera (OSIRIS-WAC, ~11° × 11° FOV, ~100 × 100 μrad/pixel), operating in the 250 nm – 1000 nm (12 filters with a 40 mm bandwidth) and 240 nm – 720 nm (14 filters with a typical bandwidth of 5 nm) spectral range, respectively (Keller et al., 2007). The NAC camera was designed to obtain high resolution images with filters optimized for the mineralogical studies of the nucleus, while the WAC camera's filters were devoted to the study of gaseous species in the comet's coma (Fornasier et al., 2015). In summary, WAC was responsible for continuous monitoring of the entire nucleus and its surroundings, while NAC focused on studying surface details. Both cameras used a 2048 × 2048-pixel backside illuminated CCD detector with a UV optimized anti-reflection coating (Tubiana et al., 2015). Although it was not a part of the scientific Rosetta payload, the NavCam (Navigational Camera) acquired images with a pixel scale ranging from ~0.7 to ~3 m, at distances from ~8 to 30 km (Basilevsky et al., 2017). The NavCam had a 5° × 5° FOV, with a 1024 x 1024 CCD sensor (Geiger et al., 2016). The NavCam was intended for optical navigation in the vicinity of the comet. However, to allow the use of the data for quantitative scientific work, Statella and Geiger (2017) performed a cross-calibration of the camera based on images from OSIRIS NAC. In Figure 5, we show images acquired by the OSIRIS system at a target distance of ~110 km, approximating the reconnaissance position planned for ASTER. At this position the swatch of NAC (A) was ~4 km × 4 km with a spatial resolution of ~2 m/pixel and WAC covered a ~20 km × 20 km with a 10 m/pixel resolution.

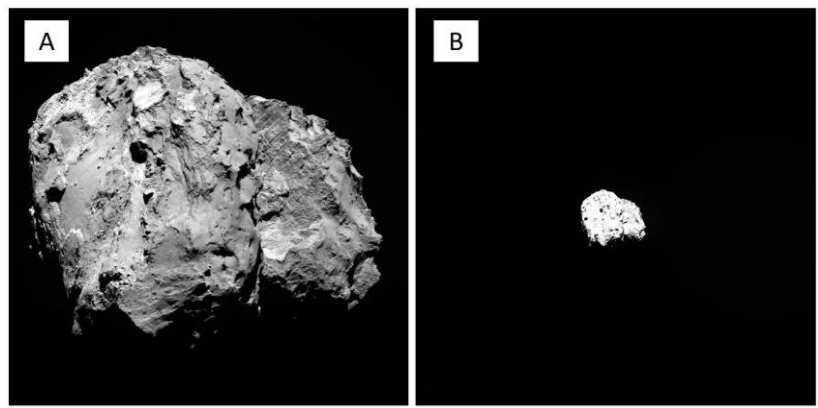

**Figure 5.** OSIRIS images taken at ~110 km from 67P comet center: (A) narrow angle, orange filter, exposure: 0.14 s; (B) wide angle, UV filter, exposure: 89 s. Time of acquisition for both: 2014-08-06, T06:44:25. For information on the filters, please refer to Keller et al. (2007). Original images credits: ESA/Rosetta/MPS for OSIRIS team MPS/UPD/LAM/IAA/SSO/INTA/UPM/DASP/IDA.

In Figure 6, we show a pair of NAC (A) and WAC (B) images taken at ~30 km from the target center, the same position planned as the mapping position for ASTER, from which it would fulfill the primary science goals. In such case, the footprint for NAC was ~1.2 km with ~0.6 m/pixels whereas the WAC footprint was ~6 km with ~3 m/pixel. It is worth noting that the 67P is larger than the 2001 SN263.

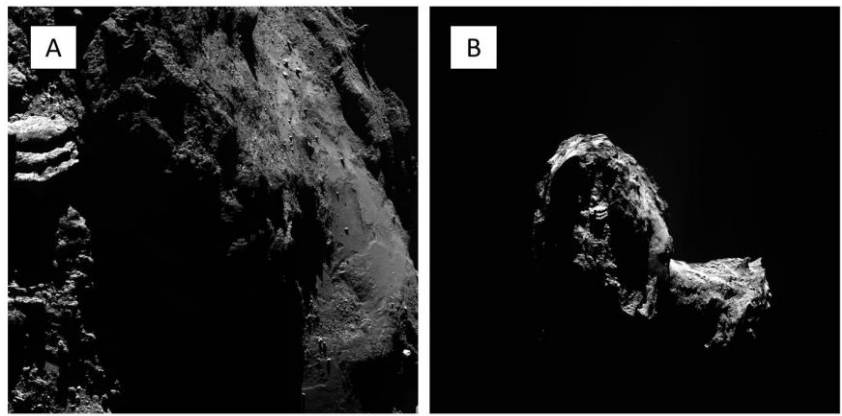

**Figure 6.** OSIRIS images taken at ~30 km from 67P comet center: (A) narrow angle, Hydra filter, exposure: 0.9 s; (B) wide angle, VIS filter, exposure: 0.65 s. Time of acquisition: 2014-11-22, T04:46:30 for NAC and 2014-11-22, T04:45:44 for WAC. For information on the filters, please refer to Keller et al. (2007). Original images credits: ESA/Rosetta/MPS for OSIRIS team MPS/UPD/LAM/IAA/SSO/INTA/UPM/DASP/IDA.

Another space mission launched with the goal of studying a NEA was OSIRIS-REx, whose target is the asteroid Bennu (~500 m in size). The primary objective of the OSIRIS-REx mission was to return pristine carbonaceous regolith from Bennu to understand how primitive asteroids contributed to the origins of life on Earth and their role as basic building blocks in planet formation (Lauretta et al., 2017); the spacecraft was launched in September 2016 and the rendezvous with the target occurred in 2018. The space mission discovered that Bennu has a negative spectral slope, typical of the B-type asteroids, and its surface is covered by fine particles (Fornasier et al., 2020). OSIRIS-REx carried a suite of cameras: the PolyCam (FOV ~0.8° × 0.8°, IFOV ~13.5 × 13.5 μrad) used mainly for very high resolution of the sample-site, MapCam (FOV ~4° × 4°, IFOV ~68 × 68 μrad) with a coarser resolution and SamCam (IFOV ~354 × 354 μrad), a wide-angle camera. Spectral resolution of the cameras ranges from 60 nm to 1000 nm, matching the *b*, *v*, *w*, and *x* ECAS filters, plus a broad-band filter in the 500 nm - 800 nm range (Lauretta et al., 2017). The Bennu sample underwent preliminary analysis through scanning electron microscopy, infrared measurements, X-ray diffraction, and chemical element analysis, which provided evidence of carbon-rich material and abundant presence of water-bearing clay minerals (Donaldson, 2023).

In September 2022, NASA's Double Asteroid Redirection Test (DART) mission, the world's first planetary defense test against NEOs, collided with Dimorphos, the 150 m moon of the Didymos asteroid system (Rivkin & Cheng, 2023). A secondary spacecraft, Light Italian CubeSat for Imaging of Asteroids (LICIACube), carried two optical cameras with monochromatic and RGB digital sensors: LUKE, with a 5.16° × 5.16° FOV, and LEIA, with a 2.07° × 2.07° FOV (Scarpa et al., 2023).

All the space missions to NEOs carried at least one multispectral camera in the payload, for such an instrument is vital for navigation, size and shape definition, age estimations of the surface, and for general mapping.

## 4. Methods

This section outlines the methodologies employed in the study. We utilized the shape model of the objects to generate a 3D model of the system, and we simulated different satellite positions to capture images of Alpha, Beta, and Gamma using the WAC and NAC cameras.

We have obtained the shape model of Alpha, Beta and Gamma from Becker et al. (2015). Each model has 2292 triangular faces and 1148 vertices. Doppler-delay radar imaging is one of the methods that can provide valuable information of the asteroid's shape and, more importantly, allow for the model to be scaled, but shape models obtained by spacecraft in situ measurements represent the ideal case (Santana-Ros et al., 2017). From the mesh grid of the objects, we used the Persistence of Vision Pty. Ltd. (2004) software POV-Ray to represent the shapes in 3D models as well as their orbits around the Sun.

For the rotation and orbital periods of Alpha, Beta and Gamma we used parameters given by Becker et al. (2015). The revolution time for Alpha was obtained from the Small-Body Database Lookup (JPL-NASA, 2023). In our experiments we adopted the revolution time of the satellite (and the camera) as the same as Alpha's and the observation time is given in hours. The orbits of Alpha in relation to the Sun and Beta in relation to Alpha were considered parallel, and for the Gamma we tilted the orbit in 15° to agree with Becker et al. (2015).

We calculated the velocities of revolution for Alpha, Beta, Gamma and the camera, and the velocity of rotation only for the bodies. Based on Becker et al. (2015), we also included the distances of Beta and Gamma from Alpha, from Alpha to the Sun, the eccentricities of the bodies, and calculated the semi-major and semi-minor radii for the orbits of the objects. We can look at the objects from different angles by changing their coordinates in the tridimensional *XYZ* planes. Figure 7 illustrates the tridimensional planes in POV-Ray.

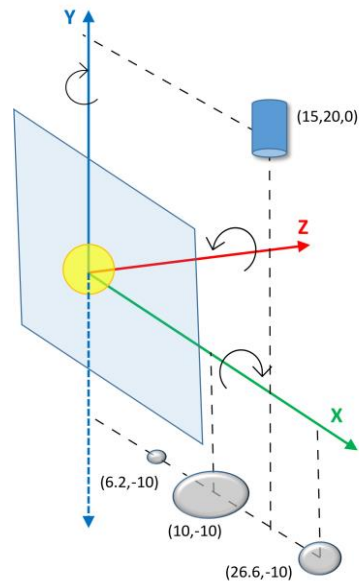

**Figure 7.** Planes follow the left-handed coordinate system in POV-Ray.

We simulated a 10º phase angle (the angle formed by the Sun-target-spacecraft vectors), so we could obtain a good view of each body of the system, without any shadows casted by the probe. Phase angle is the angle formed between the light source (such as the Sun) and an observer, as seen from a celestial object, such as a planet, a moon, or an asteroid. A low phase angle is optimal for spectroscopic investigations, while high phase angles (such as 90°) are better for morphologic studies (Bell III et al., 2002). When the phase angle is relatively small (less than 20°), the celestial body is almost fully illuminated from the spacecraft's perspective, reducing significant shadowed areas on the body. The closer the phase angle is to zero, the brighter the body appears and the smaller the shadow cast on it, as a zero sun-phase angle occurs when the Sun is directly behind the spacecraft along the approach line toward the body (Nesnas et al., 2021).

Figure 8 summarizes the main steps followed for generating the images.

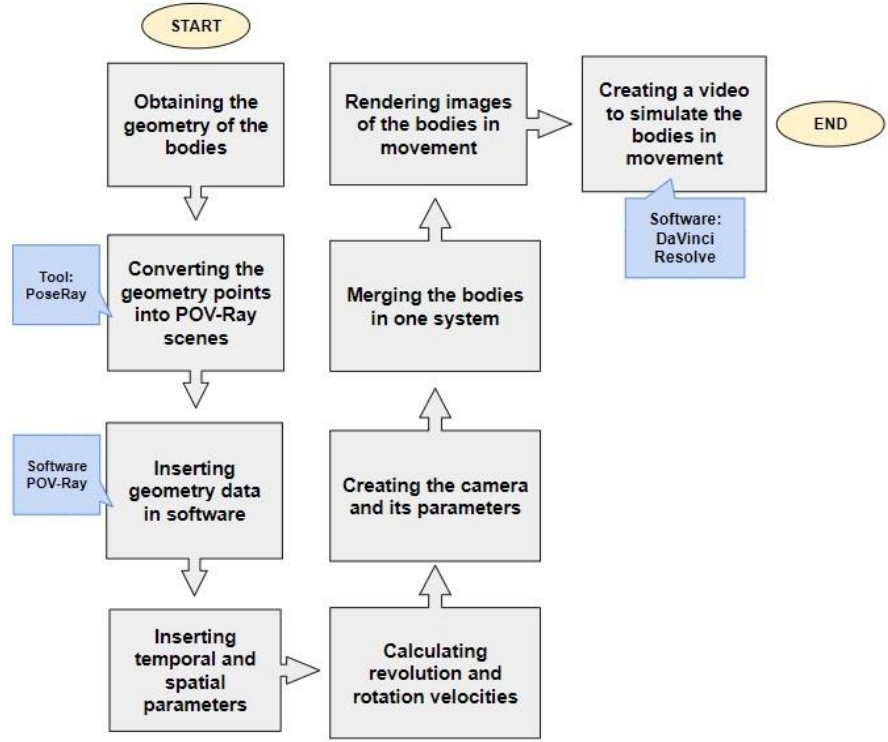

**Figure 8.** Flowchart of the main steps followed.

## 5. Results and Discussion

Amongst the main ASTER mission scientific goals, we emphasize the size and shape definition, the complete mapping of the triple system Alpha, Beta and Gamma and the age estimation of the system. For that purpose, we propose as one of the payloads for the spacecraft a multispectral camera system: one narrow-angle and another wide-angle camera, which we are calling AMC-NAC (Aster Multi-band Cameras - Narrow Angle Camera) and WAC (Aster Multi-band Cameras - Wide Angle Camera), whose specifications were devised to meet the initial condition that the Alpha body should be entirely framed at the distance of 30 km. To do so, the cameras must have a 2048 × 2048 CCD plane with a 7° × 7° FOV (~3.7 km at 30 km) and 58 × 58 μrad/pixel (~2 m/pixel at 30 km). With this configuration, the Alpha body utilizes ~1760 x 1760 pixels, Beta ~650 x 650 pixels and Gamma ~240 x 240 pixels. The depth of each pixel is proposed to be 12 bits to detect small brightness variation on the surface of the targets. The focal ratio is suggested to be set at 8, which is a compromise between speed and depth of view.

We choose the mapping position to be at 30 km from Alpha because the external region of the system is stable from ~27 to 80 km, for a wide range of inclinations, as reported by Araujo et al. (2015). Brum et al. (2021) also considered the distance of 30 km from Alpha for the ALR laser pulses. Investigating the orbital changes within the system requires the measurement of the positions between two objects, that must be framed together. By determining the relative movement of the asteroids and also their position in relation to the stars background, it is possible to calculate their orbits via astrometry and infer their masses and densities by merging the data with information provided by the laser altimeter and the spectrometer. Thus, we choose the distance of 100 km from Alpha as the reconnaissance position, but WAC will be able to obtain images from the objects at greater distances (> 300 km), enough to capture the entire system together.

We assume the spacecraft's position according to Brum et al. (2021), who designated a one-month exploration period within a five-month exploration window and defined the spacecraft's positioning between the Sun and the asteroid, back to the Sun, to get a fully illuminated body. The spacecraft trajectory is heliocentric, alongside the Alpha body.

Both the WAC and NAC cameras have identical systems, except for the focal length, which reduces costs in construction, and we will be able to obtain different images from the same distance, instead of varying the distance from the target. NAC and WAC are complementary systems. NAC is a high spatial resolution camera capable of monitoring the asteroid from relatively great distances and providing small-scale mapping of features when the spacecraft is in its mapping position (30 km), resolving craters down to 1 m of diameter, and supporting composition studies from its seven spectral filters. The NAC data will provide a detailed view of the system, surface age estimation and boulder counting. WAC has lower spatial resolution but, accordingly, a wider field-of-view. This will provide a synoptic view of the asteroid at the mapping position, allowing for body shape estimations and rotation monitoring.

In summary, WAC will produce long-term monitoring of the entire asteroid at the mapping position, so the camera must be able to frame the Alpha body completely in one single shot, whereas NAC will produce detailed imagery of smaller features with a spatial resolution of 0.25 m/pixel.

The cameras do not have native stereo capabilities but the motion of the spacecraft relative to the asteroid, coupling with a longitudinal superposition of at least 60% of the images, can be used for producing stereo-pairs, which will aid shape modeling. The camera system will also allow for surface photometry, which can give us information of the surface roughness through the applications of, for example, Hapke modeling (Hapke, 2012). These specifications should be sufficient to attend the scientific objectives of estimating the ages of the targets and their sizes and shapes. Table 2 summarizes the specifications for the camera system.

**Table 2.** Specifications for the AMC instrument on board the ASTER mission.

| Camera specifications | | |
|---|---|---|
| | **NAC** | **WAC** |
| **Sensor type** | CCD | CCD |
| **Sensor size (pixel)** | 2048 x 2048 | 2048 x 2048 |
| **Full-well (e$^-$/pxl)** | 100000 | 100000 |
| **Readout noise (e$^-$)** | 5 | 5 |
| **Dynamic range (e$^-$)** | 20000 | 20000 |
| **Gain (e$^-$/DN)** | 5 | 5 |
| **S/N at the target** | >200 | >200 |
| **Quantization (bit)** | 12 | 12 |
| **Focal ratio** | 3.4 | 3.4 |
| **FOV (°)** **Elevation and azimuth** | 0.98 (1700 m at 100 km) (512 m at 30 km) | 7.6 (13300 m at 100 km) (4000 m at 30 km) |
| **IFOV ($\mu$rad)** | 8.34 (0.85 m at 100 km) (0.25 m at 30 km) | 65 (6.5 m at 100 km) (2 m at 30 km) |
| **Alpha size (pixel)** | 3840 at 100 km 12800 at 30 km | 491 at 100 km 1638 at 30 km |
| **Beta size (pixel)** | 1440 at 100 km 4800 at 30 km | 184 at 100 km 614 at 30 km |
| **Gamma size (pixel)** | 852 at 100 km 2840 at 30 km | 109 at 100 km 363 at 30 km |

**Table 3.** Filter specifications for the AMC instrument.

| NAC and WAC specifications | |
|---|---|
| **Filters (nm)** | 235-415 (*u*) |
| | 355-560 (*b*) |
| | 495-655 (*v*) |
| | 620-835 (*w*) |
| | 775-955 (*x*) |
| | 855-1055 (*p*) |
| | 955-1155 (*z*) |
| | 500-700 (*pan*) |
| **FWHM of filters (nm)** | 260-390 (*u*) |
| | 380-495 (*b*) |
| | 520-580 (*v*) |
| | 675-745 (*w*) |
| | 812-900 (*x*) |
| | 910-1000 (*p*) |
| | 1010-1110 (*z*) |
| | 500-700 (*pan*) |
| **Effective wavelength (nm)** | 325 (*u*) |
| | 441 (*b*) |
| | 553 (*v*) |
| | 711 (*w*) |
| | 859 (*x*) |
| | 956 (*p*) |
| | 1059 (*z*) |

Regarding the focus, adopting a $f\#$ of 3.4, like the one used by the NEAR Shoemaker mission (MSI-NEAR Multispectral Imaging System), and considering a circle of confusion (cc) of 0.029 and a focal length of ~180 mm, the hyperfocal distance is estimated to be ~330 m. Therefore, the focal range must be the interval [~160 m, infinity].

For the ASTER mission, the NAC camera angle is 0.98° × 0.98° and the WAC camera angle is 7.6° × 7.6°. If the NAC camera angle were narrower, it would require capturing more images of each body, leading to increased memory consumption. If we adopted a wider WAC camera angle, we could lose detail.

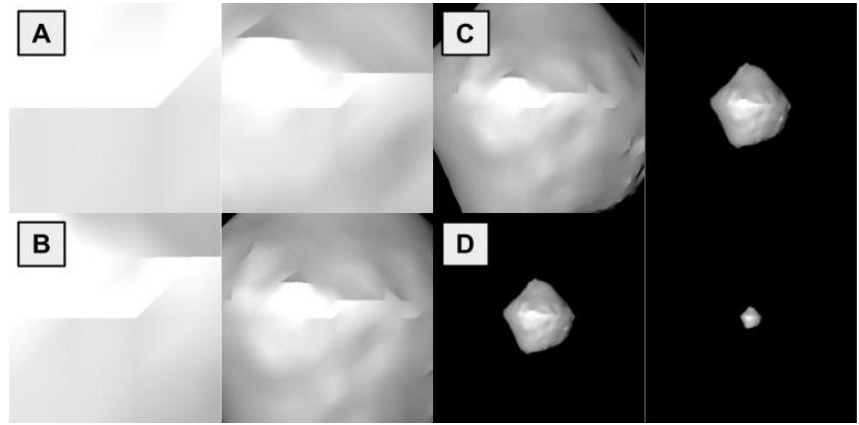

**Figure 9.** Viewing of Alpha from different FOV angles, at 30 km and 100 km from Alpha, respectively: 0.5° (A), NAC=0.98° (B), WAC=7.6° (C) and 15° (D).

Based on the 2001 SN263 spectra, and the heritage from previous missions, we propose to use a set of 7 ECAS filters, *u*, *b*, *v*, *w*, *x*, *p* and *z* (Figure 10), similar to the ones used by AMICA in the Hayabusa mission, plus one broad-band filter for navigation (500 nm to 700 nm), and a quantization in 12 bits. Utilizing these filters, WAC will be able to fully map the three bodies in different spectral bands, revealing potential variations in composition and possibly a differentiation between the system's components. The NAC images, with higher resolution, will be combined to create thematic maps (spectral classes) of surface mineralogy. We opt for extrapolation into the near-infrared range due to the laser altimeter and spectrometer's operation in this spectrum. This ensures their coincident focal points, enabling mutual calibration throughout the mission and utilizing the camera-generated coordinate systems for integrating measurements from the other two instruments.

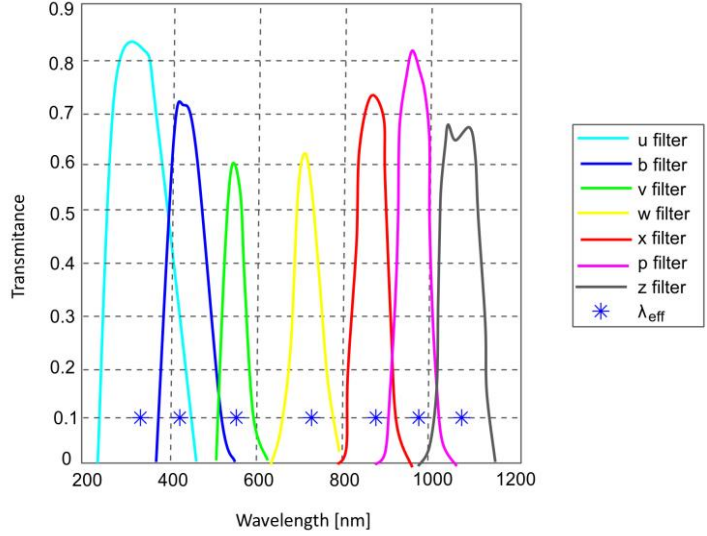

**Figure 10.** ECAS filters transmission curves as proposed by Tedesco et al. (1982).

In Figure 10, the effective wavelength $\lambda_{eff}^i$ for each filter *i* was calculated as:

$$\lambda_{eff}^i = \frac{\int_{\lambda 1}^{\lambda 2} \lambda \cdot \tau_i(\lambda) \cdot d\lambda}{\int_{\lambda 1}^{\lambda 2} \tau_i(\lambda) \cdot d\lambda} \tag{1}$$

Where $\tau_i(\lambda)$ is the spectral transmittance for the filter *i*.

To cover the bodies with 10% latitudinal and 60% longitudinal superposition, the WAC camera requires seven images (for one filter) each for Alpha, Beta, and Gamma at distances of 100 km and 30 km. Each set of seven images consumes 0.054 GB of memory (uncompressed data), resulting in a total memory usage of 0.324 GB per filter. The NAC camera requires, for Alpha, 43 images at 100 km (0.34 GB) and 405 images at 30 km (3.2 GB), for Beta, seven images at 100 km (0.054 GB) and 64 images at 30 km (0.5 GB), for Gamma, seven images at 100 km (0.054 GB) and 24 images at 30 km (0.19 GB), resulting in a total memory usage of ~4 GB per filter. The minimum time required to capture all the images is estimated to be 33 hours and 14 minutes. Additional time is necessary for capturing images of Gamma during the period either hidden or partially hidden by Alpha, around two hours or 11% of Gamma's rotation period.

We have performed simulations of the NAC and WAC views using the asteroid models shown in Figure 1 and the software PovRay. The cameras were modeled accordingly with their specifications given in Table 2. In Figure 11 we show the views of the WAC and NAC when pointing at the Alpha body. Panels 11(A) and 11(B) simulate the WAC viewing from 100 km and 30 km, respectively. In panels 11(C) and 11(D) the NAC is placed at 100 km and 30 km, respectively, from Alpha. For these simulations, the spacecraft is following Alpha on its orbital plane during the asteroid revolution around the Sun. In Figures 12 and 13 the cameras point at Beta and Gamma, respectively, while following the Alpha movement.

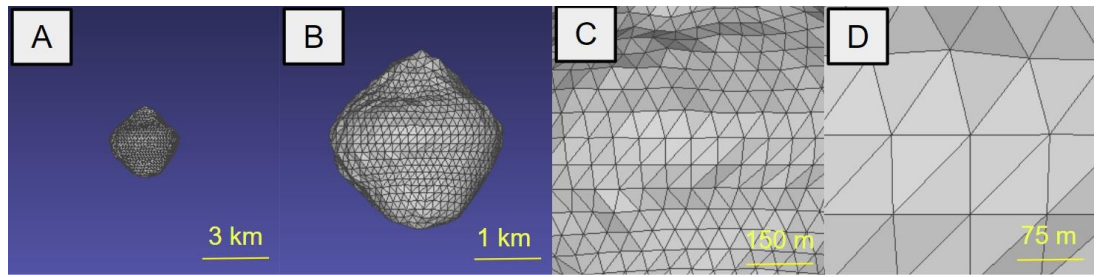

**Figure 11.** Viewing of WAC and NAC with both pointing at Alpha. In (A) and (B): WAC viewing at 100 km and 30 km, respectively, from Alpha; In (C) and (D): NAC viewing at 100 km and 30 km, respectively, from Alpha.

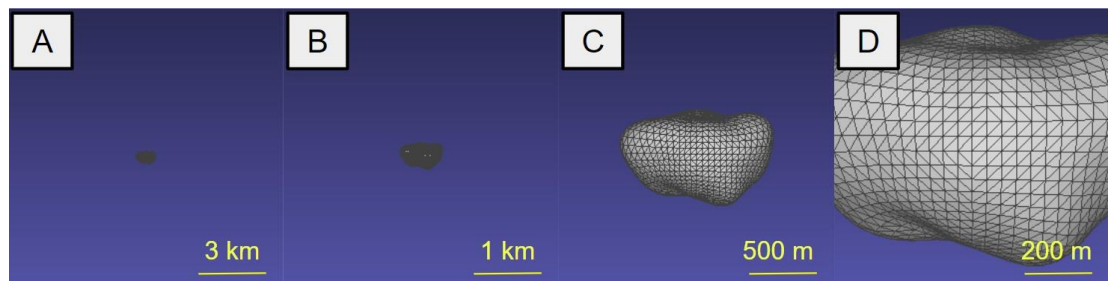

**Figure 12.** Viewing of WAC and NAC with both pointing at Beta. In (A) and (B): WAC viewing at 100 km and 30 km, respectively, from Alpha; In (C) and (D): NAC viewing at 100 km and 30 km, respectively, from Alpha.

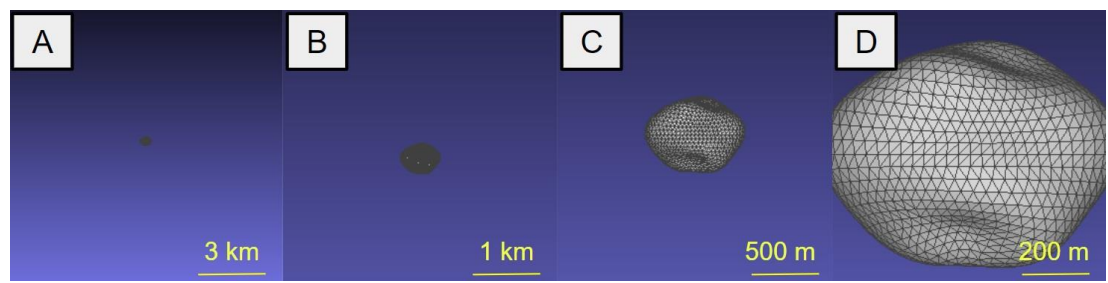

**Figure 13.** Viewing of WAC and NAC with both pointing at Gamma. In (A) and (B): WAC viewing at 100 km and 30 km, respectively, from Alpha; In (C) and (D): NAC viewing at 100 km and 30 km, respectively, from Alpha.

The wide-angle camera was chosen because it can provide a wide view of the system, while the narrow angle camera gives more detail of the objects Alpha, Beta and Gamma. We simulated images with MeshLab software to show the wireframe model of the asteroid and the current level of detail we have of the system. More simulations are given in Figures 14-16.

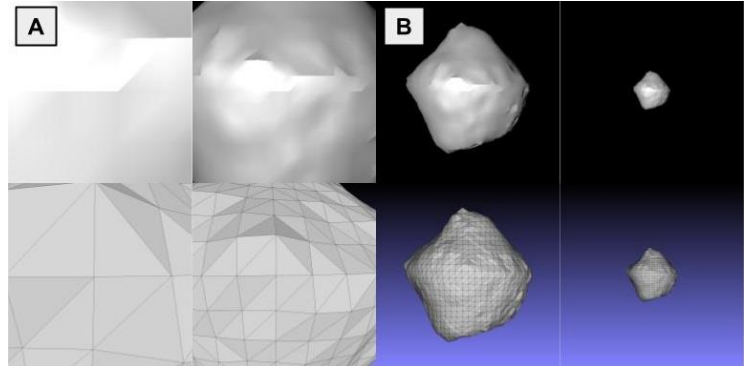

**Figure 14.** Simulation of the asteroid area that will be seen in detail by the NAC and WAC cameras, considering the current wireframe model. In (A): NAC, POV-Ray images (top) and MeshLab images (bottom); In (B): WAC, POV-Ray images (top) and MeshLab images (bottom).

The choice of 10° for the phase angle is because long term variations in apparent brightness of an asteroid depend mainly on its distance to the Sun and to the observer, and the phase angle (Santana-Ros et al., 2017). In a study realized by Jawin et al. (2022), the authors created mosaics of the asteroid Bennu, using a 30° phase angle to remove illumination-dependent effects.

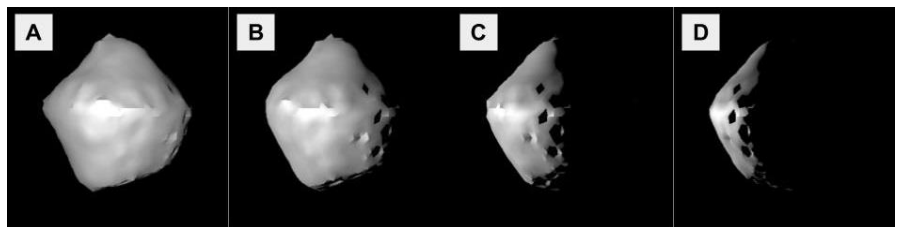

**Figure 15**. Viewing of Alpha from different phase angles: 10° (A), 45° (B), 90° (C) and 120° (D).

According to Fang et al. (2011) and Becker et al. (2015), the Gamma's spin pole could have a ~15° inclination. We opted to rotate Gamma to simulate its inclined orbit inside the parameters of the system, to make sure it was in the right place. Additionally, Gamma is tidally locked (Becker et al., 2015), meaning that we can only observe one phase of it from the same perspective, as its rotation period matches its orbital period. We simulated an elevation in the satellite's orbit to assess the feasibility of visualizing Gamma behind Alpha. While most of the displayed section of Gamma corresponds to its pole and part of the object is obscured, it is possible to capture images of Gamma during the period when Alpha is positioned in front of it.

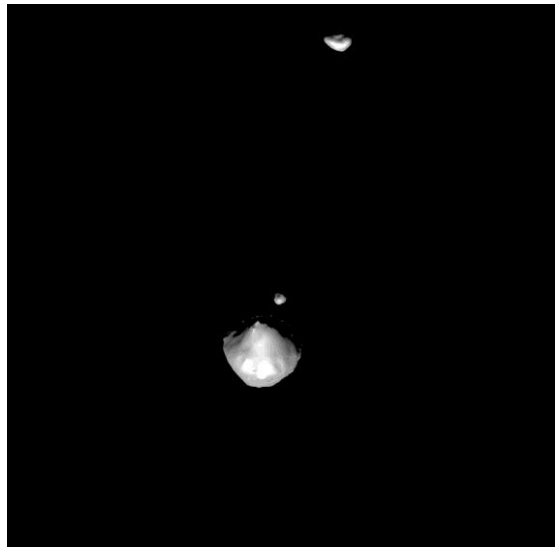

**Figure 16**. Simulated observation of Gamma positioned behind Alpha from a potential elevated orbit of the satellite. Beta is also visible at the top of the image.

6. **Final remarks**

This article has systematically defined the camera system parameters of the ASTER mission, simulated the expected images that would be obtained by the cameras and indicated their constraints, as presented in Sections 4 and 5.

The simulations generated from the multispectral camera system parameters show the possible images that should be captured by the cameras in the reconnaissance position (100 km) and in the mapping position (30 km). We demonstrate that Alpha is entirely visible only in the WAC images, while the NAC is expected to reveal surface details. Beta seems relatively small in the WAC images, whereas we obtain a broad view from the NAC at 100 km distance. Gamma, smaller than Beta, should be imaged by the NAC, whereas the WAC images should be able to show its inclined orbit around Alpha. In addition, due to the close distance to Alpha, the Gamma object cannot be entirely mapped from the reconnaissance nor mapping orbit if the camera follows Alpha. To do so, we would have to elevate the camera's orbit to be able to see Gamma behind Alpha in its revolution movement. Regarding additional practical contributions, the method employed to simulate images generated by satellite cameras can be applied to other scenarios where the target requires imaging, extending beyond the field of planetary geology.

**Acknowledgements**

We would like to thank the UNESP's Orbital Dynamics and Planetology Group (GDOP) for providing us with the shape models of Alpha, Beta, and Gamma.

RS acknowledges support by the DFG German Research Foundation (project 446102036), Fundação de Amparo à Pesquisa do Estado de São Paulo (FAPESP) – Proc. 2016/24561-0.